\documentclass[11pt]{amsart}
\usepackage{amsmath}
\usepackage{amsfonts,amssymb,latexsym}
\usepackage{amscd}
\usepackage{mathrsfs}
\usepackage{euscript}
\usepackage{amsthm}
\newtheorem{theorem}{\bf{Theorem}}[section] 
\newtheorem{corollary}[theorem]{\bf{Corollary}} 

\newtheorem{lemma}[theorem]{\bf{Lemma}} 
\newtheorem{proposition}[theorem]{\bf{Proposition}}

\newtheorem{definition}[theorem]{\bf{Definition}}

\theoremstyle{remark}
\newtheorem*{example}{Example}

\newtheorem{remark}{Remark}
\newtheorem*{remarks}{Remarks}

\usepackage{graphics}
\numberwithin{equation}{section}
\def\Bbb{\mathbb}

\def\logo{\raisebox{-10.5\p@}{\hb@xt@85\p@{\includegraphics{gft.eps}\hfil}}}

\def\un{1\kern-3pt \rm I}
\def\ptoday{{\ifcase\month 
\or January, \or February, \or March, \or April,\or May, 
\or June, \or July, \or August, \or September, \or October, 
\or November, \or December,\fi\ \number \year}}

\input epsf
\usepackage[dvips]{graphicx}

\def\stackrel#1#2{\mathrel{\mathop{#2}\limits^{#1}}}

\def\dj{\hbox{d\kern-0.347em \vrule width 0.3em height 1.252ex depth
-1.21ex \kern 0.051em}}

\begin{document}
\title[A Note on Superdistributions and Wavefront Set]
      {\sl A Note on Superdistributions and Wavefront Set}
        
\author{Daniel H.T. Franco}
\address{Centro de Estudos de F\'\i sica Te\'orica, Setor de F\'\i sica--Matem\'atica\\
         Rua Rio Grande do Norte 1053/302, Funcion\'arios \\
         Belo Horizonte, Minas Gerais, Brasil, CEP:30130-131.}
\email{dhtf@terra.com.br}

\date{August 17, 2006}
\begin{abstract}
We present a simple and new method of constructing superdistributions on superspace
over a Grassmann-Banach algebra, which close to the de Rham's ``currents'' defined as
dual objects to differential forms. The paper also contains the extension of the
H\"ormander's description of the singularity structure (wavefront set) of a distribution
to include the supersymmetric case.
\end{abstract}

\maketitle


\section{Introduction}
\hspace*{\parindent}
In this work, we extend the definition of the objects most widely used in physics:
distributions. The distribution theory is a generalization of the classical analysis,
which makes it possible to deal in a systematic way difficulties as the unpleasant
fact that not every function is differentiable. As a matter of fact, the space of
distributions is essentially the smallest extension of the space of continuous
functions where differentiation is always well defined. The theory was intensively
developed by many mathematicians and theoretical physicists, mainly in connection
with the needs of theoretical and mathematical physics. It one relies fundamentally
on the notion of topological vector spaces. The Quantum Field Theory (QFT) perhaps is the
example more important where technical results from distribution theory are required.
In particular, Schwartz's theory of tempered distributions became fundamental to the
G\r{a}rding-Wigthman axiomatization of relativistic QFT~\cite{SW,BLOT}. In the same way
the Fourier analysis of distributions plays an important role in the QFT, mainly
in the spectral analysis of singularities. With this paper, we intend to define
superdistributions and Fourier transformations in the supersymmetric field theory in
the spirit of Schwartz's distributions and in the spirit of H\"ormander's spectral
analysis of singularities of distributions.

Over the last decades, supersymmetric quantum theories
have been studied intensively with the belief that such theories may play a part in a
unified theory of fundamental forces, and many issues are understood much better now.
These theories are usually characterized by their invariance properties with respect
to transformations that involve anticommuting parameters. The latter play an essential
role in the formulation of supersymmetric theories and their use sometimes facilities
calculations, for instance in perturbation theory. As it occurs with the ordinary quantum
field theories, supersymmetric field theories are also deeply connected to the presence
of ultraviolet divergences, in a naive approach. However, physicists have soon learned
how to make sense out of them in a mathematically proper way through the procedure now
known as renormalization (a comprehensive account of the quantum theory through
the algebraic renormalization approach can be found in the textbook by Piguet
and Sibold~\cite{PiSi}). As first indicated by Wess and Zumino, supersymmetry is preserved
by renormalization and further leads to a less divergent than conventional field theoretic
model.

It is already well-known that the singularity structure of Feynman (or more precisely
Wightman) superfunctions is completely associated with the ``bosonic'' sector of the
superspace -- the body of superspace. This result can be mainly justified by the heuristic
form of defining superspace and superfields. It is, therefore, a natural question to ask how
a mathematically rigorous definition of the structure of these singularities can be given. 
Although claims exist that such a result is completely obvious, we do not think that a clear
proof is available in the published literature, to the best of our knowledge. However,
to our great surprise, such a proof does exist and is extremely simple. The key ingredients
in our analysis are the notion of the wavefront set~\cite{Hor2}--\cite{DH} of a
superdistribution and the appropriate construction of Rogers of a superspace and
superfields~\cite{Rogers}. The notion of wavefront set was introduced by the mathematicians
H\"ormander and Duistermaat~\cite{Hor1,DH} in the seventies and it is growing of importance,
with a range of applications going beyond the original problems of linear partial equations.
It has received, in the last years, a lot of attention from community of theoretical
physicists in order to solve some important problems, such as the characterization of
the spectral condition for a QFT on a general manifold~\cite{Rad,BFK}.

This note is organized as follow: in Sec. 2, for the convenience of readers,
we shall briefly review some few basic properties
of superspaces based on the Rogers' work~\cite{Rogers}. In Sec. 3, a new formulation of
superdistributions on superspace is presented.\footnote{An alternative formulation of
superdistributions is given in the Ref.~\cite{NK}.}
Such a formulation close to the de Rham's ``currents''
defined as dual objects to differential forms~\cite{Rham}. In Sec. 4, we extend the
notion of the wavefront set of a superdistribution. The well-known result that the
singularities of a superdistribution may be expressed in a very simple way through the
ordinary distribution is proved by functional analytical methods. Sec. 5 contains
our conclusions. Finally, for sake of completeness, in the Appendix A we recall some
properties of the microlocal analysis.

\section{Notions of Superspace}
\label{NoSmf}
This section introduces some few basic fundamentals on the theory of
superspace. We follow here the work of Rogers~\cite{Rogers}. 
Rogers' theory has an advantage, a superspace is an ordinary Banach manifold endowed with a
Grassmann algebra structure, so that the topological constructions have their standard meanings.

We start by introducing first some definitions and concepts of a Grassmann-Banach
algebra, i.e., a Grassmann algebra endowed with a Banach algebra structure.
Let $L$ be a finite positive integer. Denote by ${\mathscr G}$
a Grassmann algebra, such that ${\mathscr G}$ can naturally be decomposed
as the direct sum ${\mathscr G}={\mathscr G}_0 \oplus {\mathscr G}_1$, where
${\mathscr G}_0$ consists of the even (commuting) elements and ${\mathscr G}_1$
consists of the odd (anti-commuting) elements in ${\mathscr G}$, respectively.
Let $M_L$ denote the set of sequences 
$\{(\mu_1,\ldots,\mu_k)\mid
1 \leq k \leq L; \mu_i \in {\Bbb N}; 1 \leq \mu_1 < \cdots < \mu_k \leq L\}$.
Let $\Omega$ represent the empty sequence in $M_L$, and $(j)$ denote the sequence
with just one e\-le\-ment $j$. A basis of ${\mathscr G}$ is given by  monomials
of the form $\{\xi_\Omega,\xi^{\mu_1}\xi^{\mu_2},\ldots,\xi^{\mu_1} 
\xi^{\mu_2} \cdots \xi^{\mu_k}\}$ for all $\mu \in M_L$, such that 
$\xi_{\Omega}={\un}$ and 
$\xi^{(i)}\xi^{(j)}+\xi^{(j)}\xi^{(i)}=0$ for 
$1\leq i,j \leq L$. Futhermore, there is no other independent relations 
among the generators. By ${\mathscr G}_L$ we 
denote the Grassmann algebra with $L$ generators, where the even and 
the odd elements, respectively, take their values. $L$ being assumed a finite 
integer (the number of generators $L$ could be possibly infinite),
it means that the sequence terminates at $\xi^{1}\ldots \xi^{L}$ and there are
only $2^L$ distinct basis elements. An arbitrary element $q \in {\mathscr G}_L$
has the form
\begin{equation}
q=q_{\bf b} + \sum_{(\mu_1,\ldots,\mu_k)\in M_L}
q_{\mu_1,\ldots,\mu_k}\xi^{\mu_1}\cdots \xi^{\mu_k}\,\,,
\end{equation}
where $q_{\bf b},q_{{\mu_1\ldots \mu_k}}$ are real numbers. An even or odd element is
specified by $2^{L-1}$ real parameters. The number $q_{\bf b}$ is called the body of $q$, while
the remainder $q-q_{\bf b}$ is the soul of $q$, denoted $s(q)$. The element $q$ is invertible
if, and only if, its body is non-zero.

With reference to supersymmetric field theories, the commuting variable $x$ has
the form
\begin{equation}
x=x_{\bf b}+x_{ij}\xi^i \xi^j+x_{ijkl}\xi^i \xi^j \xi^k \xi^l+\cdots\,\,,
\label{snx}
\end{equation}
where $x_{\bf b},x_{ij},x_{ijkl},\ldots$ are real variables.
Similarly, the anticommuting variables (in the Weyl representation) 
$\theta$ and $\bar{\theta}=(\theta)^*$ have the form
\begin{equation}
\theta=\theta_{i}\xi^i+\theta_{ijk}\xi^i \xi^j \xi^k +\cdots\,\,,
\quad
\bar{\theta}=\bar{\theta}_{i}\xi^i+
\bar{\theta}_{ijk}\xi^i \xi^j \xi^k +\cdots\,\,,
\label{snt}
\end{equation}
where $\theta_{i},\theta_{ijk},\ldots$ are complex variables.
The summation over repeated indices is to be understood unless
otherwise stated. As pointed out by Vladimirov-Volovich~\cite{Vlavolo}, 
from the physical point of view, superfields are not functions of 
$\theta_{i},\theta_{ijk},\ldots$ and $x_{\bf b},x_{ij},x_{ijkl},\ldots$, 
but only depend on these variables through $\theta$ and $x$,
as it occurs with ordinary complex analysis where analytic functions 
of the complex variables $z=x+iy$ are not arbitrary functions of the 
variables $x$ and $y$, but functions that depend on $x$ and $y$ 
through $z$.

The Grassmann algebra may be topologized. Consider the complete norm 
on ${\mathscr G}_L$ defined by~\cite{Rudo}: 
\begin{equation}
\|q\|_p=\left(|q_{\bf b}|^{p}+\sum_{(\mu)=1}^L
|q_{{\mu_1 \ldots \mu_k}}|^p\right)^{1/p}\,\,.
\label{norma}
\end{equation}
A useful topology on ${\mathscr G}$ is the
topology induced by this norm. The norm $\|\cdot\|_1$ is called the Rogers
norm and ${\mathscr G}_L(1)$ the Rogers algebra~\cite{Rogers}.
The Grassmann algebra ${\mathscr G}$ equipped with the norm (\ref{norma})
becomes a Banach space. In fact ${\mathscr G}$ becomes a
Banach algebra, i.e., $\|{\un}\|=1$ and $\|qq^\prime\| \leq \|q\|\|q^\prime\|$
for all $q,q^\prime \in {\mathscr G}$.

\begin{definition}
A Grassmann-Banach algebra is
a Grassmann algebra endowed with a Banach algebra structure.  
\end{definition}

A superspace must be constructed using as a building block a
Grassmann-Banach algebra ${\mathscr G}_L$ and not only a
Grassmann algebra.

\begin{definition}
Let
${\mathscr G}_L={\mathscr G}_{L,0} \oplus {\mathscr G}_{L,1}$ be
a Grassmann-Banach algebra. Then the $(m,n)$-dimensional superspace
is the topological space 
${\mathscr G}_L^{m,n}={\mathscr G}_{L,0}^m \times {\mathscr G}_{L,1}^n$,
which generalizes the space ${\Bbb R}^m$,
consisting of the Cartesian product of $m$ copies of the even 
part of ${\mathscr G}_L$ and $n$ copies of the odd part.
\end{definition}

In supersymmetric quantum field theory, superfields are functions in 
superspace usually given by their (terminating) standard expansions 
in powers of the odd coordinates
 \begin{equation}
F(x,\theta,\bar{\theta})=\sum_{(\gamma)=0}^\Gamma f_{(\gamma)}(x)
(\theta)^{(\gamma)}\,\,,
\label{spfield} 
\end{equation}
where $(\theta)^{(\gamma)}$ comprises all monomials in the anticommuting
variables $\theta$ and $\bar{\theta}$ (belonging to odd part of 
a Grassmann-Banach algebra) of degree $|\gamma|$; $f_{(\gamma)}(x)$
is called a component field, whose Lorentz properties are determined by
those of $F(x,\theta,\bar{\theta})$ and by the power $(\gamma)$ of $(\theta)$.
The following notation, extended to more than one $\theta$ variable,
is used (\ref{spfield}): 
$(\theta)=(\theta_1,\bar{\theta}_1,\ldots,\theta_n,\bar{\theta}_n)$, and
$(\gamma)$ is a multi-index 
$(\gamma_1,\bar{\gamma}_1,\ldots,\gamma_n,\bar{\gamma}_n)$ with
$|\gamma|=\sum_{r=1}^n(\gamma_r+\bar{\gamma}_r)$ and $(\theta)^{(\gamma)}=\prod_{r=1}^m 
\theta_r^{\gamma_r}\bar{\theta}_r^{\bar{\gamma}_r}$. 
In Eq.(\ref{spfield}), for a (4,4)-dimensional superspace, $\Gamma=(2,2)$.

Rogers~\cite{Rogers} considered superfields in ${\mathscr G}_L^{m,n}$ 
as $G^\infty$ superfunctions, i.e., functions whose coefficients $f_{(\gamma)}(x)$
of their expansions are smooth functions of ${\Bbb R}^m$ into ${\mathscr G}_L$,
extended from ${\Bbb R}^m$ to all of ${\mathscr G}_L^{m,0}$ by $z$-continuation.
Throughout the remainder of this paper the prefix ``super'' is used for entities
involving odd Grassmann variables.

\begin{definition}
Let $U$ be an open set in ${\mathscr G}_L^{m,0}$ and 
$\epsilon:{\mathscr G}_L^{m,0} \rightarrow {\Bbb R}^m$ be the body projection
which associates to each $m$-tuple $(x_1,\ldots,x_m)$ in ${\mathscr G}_L^{m,0}$
the $m$-tuple $(\epsilon(x_1),\ldots,\epsilon(x_m))$ in ${\Bbb R}^m$.
Let $V$ be an open set in ${\Bbb R}^m$ with $V=\epsilon(U)$.
We get through $z$-continuation -- or ``Grassmann analytic continuation'' --
of a function $f\in C^\infty(V,{\mathscr G}_L)$ a function
$z(f)\in G^\infty(U,{\mathscr G}_L)$, which admits an expansion in powers of the soul
of $x$
\begin{align*}
z(f)(x_1,\ldots,x_m)=\sum_{i_1=\cdots=i_m=0}^L \frac{1}{i_1!\cdots i_m!}
\left[\partial_1^{i_1} \cdots \partial_m^{i_m}\right]f(\epsilon(x))
s(x_1)^{i_1} \cdots s(x_m)^{i_m}\,\,,
\end{align*}
where $s(x_i)=(x_i-\epsilon(x_i))$ and $\epsilon(x_i)=(x_i)_b$.
\label{galo}
\end{definition}

One should keep always in mind that the continuation involves only
the even variables $z:C^\infty(\epsilon(U))\rightarrow G^\infty(U)$,
and that $z(f)(x_1,\ldots,x_m)$ is a supersmooth function
if their components are smooth for soulless values of $x$. 
This justifies the formal manipulations in the physics literature, 
where superfields are manipulated as if their even arguments were ordinary
numbers~\cite{Rabin}: a supersmooth function is completely determined when
its components are known on the body of superspace.

According to Definition \ref{galo}, the superfield $F(x,\theta,\bar{\theta})
\in G^\infty(U,{\mathscr G}_L)$ admits an expansion
\[
F(x,\theta,\bar{\theta})=\sum_{(\gamma)=0}^\Gamma z(f_{(\gamma)})(x)
(\theta)^{(\gamma)}\,\,,
\]
but here with suitable $f_{(\gamma)}\in C^\infty(\epsilon(U),{\mathscr G}_L)$. 

\section{Distributions on the Superspace}
\label{DiSmf}
We begin by introducing the concept of superdistributions as the dual space of supersmooth
functions in ${\mathscr G}_L^{m,0}$, with compact support, equipped with an appropriate
topology, called {\em test superfunctions}. This can be done relatively straightforward in
analogy to the notion of distributions as the dual space to the space $C_0^\infty(U)$ of
functions on an open set $U \subset {\Bbb R}^m$ which have compact support, since the
spaces ${\mathscr G}_L^{m,0}$ and ${\mathscr G}_L^{m,n}$ are regarded as ordinary vector
spaces of $2^{L-1}(m)$ and $2^{L-1}(m+n)$ dimensions, respectively, over the real numbers. 
 
Let $\Omega \subset {\Bbb R}^m$ be an open set. $\Omega=\epsilon(U)$ regarded
as a subset of ${\mathscr G}_L^{m,0}$, it is identified with the body of some
domain in superspace. Let $C_0^\infty(\Omega,{\mathscr G}_L)$ be the space of 
${\mathscr G}_L$-valued smooth functions with compact support in ${\mathscr G}_L$.
Every function $f \in C_0^\infty(\Omega,{\mathscr G}_L)$ can be expanded in terms
of the basis elements of ${\mathscr G}_L$ as:
\begin{align}
f(x)=\sum_{(\mu_1,\ldots,\mu_k)\in M_L^0}f_{\mu_1,\ldots,\mu_k}(x)
\xi^{\mu_1}\cdots \xi^{\mu_k}\,\,,
\end{align}
where $M_L^0 {\stackrel{\rm def}{=}}\{(\mu_1,\ldots,\mu_k)\mid
0 \leq k \leq L; \mu_i \in {\Bbb N}; 1 \leq \mu_1 < \cdots < \mu_k \leq L\}$
and $f_{\mu_1,\ldots,\mu_k}(x)$ is in the space $C^\infty_0(\Omega)$ of
real-valued smooth functions on $\Omega$ with compact support. Thus, it follows that
the space $C^\infty_0(\Omega,{\mathscr G}_L)$ is isomorphic to the space 
$C^\infty_0(\Omega) \otimes {\mathscr G}_L$~\cite{NK}. In accordance with the Definition
\ref{galo}, the smooth functions of $C_0^\infty(\Omega,{\mathscr G}_L)$ can be extended
from $\Omega \subset {\Bbb R}^m$ to $U \subset {\mathscr G}_L^{m,0}$ by Taylor expansion.

In order to define superdistributions, we need to give a suitable topological
structure to the space $G_0^\infty(U,{\mathscr G}_L)$ of ${\mathscr G}_L$-valued
superfunctions on an open set $U \subset {\mathscr G}_L^{m,0}$ which have compact support.
According to a proposition by Rogers, every $G^\infty$ superfunction on a
compact set $U \subset {\mathscr G}_L^{m,0}$ can be considered as a real-valued
$C^\infty$ function on $U \subset {\Bbb R}^N$, where $N=2^{L-1}(m)$, regarding 
${\mathscr G}_L^{m,0}$ and ${\mathscr G}_L$ as Banach spaces. In fact, the identification of
${\mathscr G}_L^{m,0}$ with ${\Bbb R}^{2^{L-1}(m)}$ is possible~\cite{CaReTe}.
We have here an example of functoriality. Indeed, let $X$ and $Y$ denote a
$G^\infty$ supermanifold and a Banach manifold $C^\infty$, respectively. Then
with each supermanifold $X$ we associate a Banach manifold $Y$, via a {\em covariant}
functorial relation $\lambda:X \rightarrow Y$, and with each $G^\infty$ map $\phi$ defined
on $X$, a $C^\infty$ map $\lambda(\phi)$ defined on $Y$~\cite{CaReTe}. 

Following, we shall first consider only the subset $C^\infty_K$ of
$C^\infty_0(U \subset {\Bbb R}^N)$ which consists of functions with support in a fixed
compact set $K$. Since by construction $C^\infty_K$ is a Banach space, the functions
$C^\infty_K$ have a natural topology given by the finite family of norms
\begin{align}
\|\phi\|_{K,m}=\sup_{\stackrel{|p|\leq m}{x \in K}}|D^p \phi(x)|\,\,,
\qquad D^p=\frac{\partial^{|p|}}{\partial x^{p_1}_1 \cdots \partial x^{p_m}_m}\,\,, 
\label{snorm} 
\end{align}
where $p=(p_1,p_2,\ldots,p_m)$ is a $m$-tuple of non-negative integers, and
$|p|=p_1+p_2+\ldots+p_m$ defines the order of the derivative. Next, let $U$ be considered as
a union of compact sets $K_i$ which form an increasing family $\{K_i\}_{i=1}^\infty$, such
that $K_i$ is contained in the interior of $K_{i+1}$. That such family exist follows from the
Lemma 10.1 of~\cite{Treves1}. Therefore, we think of $C^\infty_0(U \subset {\Bbb R}^N)$ as
$\bigcup_i C^\infty_{K_i}(U \subset {\Bbb R}^N)$. We take the topology of
$C^\infty_0(U \subset {\Bbb R}^N)$ to be given by the strict
inductive limit topology of the sequence $\{C^\infty_{K_i}(U \subset {\Bbb R}^N)\}$. Of 
another way, we may define convergence in $C^\infty_0(U \subset {\Bbb R}^N)$
of a sequence of functions $\{\phi_k\}$ to mean that for each $k$, one has
${\rm supp}\,\,\phi_k \subset K \subset U \subset {\Bbb R}^N$ such that for a function
$\phi \in C^\infty_0(U \subset {\Bbb R}^N)$ we have $\|\phi -\phi_k\|_{K,m}\rightarrow 0$ as
$k \rightarrow \infty$. This notion of convergence generates a topology which makes
$C^\infty_0(U \subset {\Bbb R}^N)$, certainly, a topological vector space. 

Now, let $\sf F$ and $\sf E$ be spaces of smooth functions with compact support defined on
$U \subset {\mathscr G}_L^{m,0}$ and $U \subset {\Bbb R}^N$, respectively.
If $\lambda:{\sf E}\rightarrow{\sf F}$ is a {\em contravariant} functor which associates
with each smooth function of compact support in ${\sf E}$, a smooth function of compact
support in ${\sf F}$, then we have a map
\begin{align}
\|\phi\|_{K,m} \longrightarrow \|\lambda(\phi)\|_{K,m}\,\,, 
\label{snorm0} 
\end{align}
providing $G_0^\infty(U,{\mathscr G}_L)$ with a limit topology induced by a
finite family of norms.

We now take a result by Jadczyk-Pilch~\cite{JaPi}, later refined by Hoyos {\it et
al}~\cite{HoQuRaUr}, which establishes as a natural domain of definition for supersmooth
functions a set of the form $\epsilon^{-1}(\Omega)$, where $\Omega$ is open in ${\Bbb R}^m$.
Let $\epsilon^{-1}(\Omega)$ be the domain of definition for a superfunction
$f \in G_0^\infty(\epsilon^{-1}(\Omega),{\mathscr G}_L)$, where $\epsilon^{-1}(\Omega)$ is
an open subset in ${\mathscr G}_L^{m,0}$ and $\Omega$ is an open subset in ${\Bbb R}^m$, and
let $\widetilde{\phi} \in C^\infty_0(\Omega,{\mathscr G}_L)$ denotes the restriction of
$\phi$ to $\Omega \subset {\Bbb R}^m \subset {\mathscr G}_L^{m,0}$. Then, it follows that 
$(\partial^{p_1}_1 \cdots \partial^{p_m}_m \phi)^{\widetilde{}}= \partial^{p_1}_1 
\cdots \partial^{p_m}_m{\widetilde{\phi}}$, where the derivatives on the
right-hand side are with respect to $m$ real variables. Now, suppose $\Omega=\bigcup_i
\widetilde{K}_i$ where each $\widetilde{K}_i$ is open and has compact closure in
$\widetilde{K}_{i+1}$. It follows that $C^\infty_0(\Omega,{\mathscr G}_L)=\bigcup_i
C^\infty_{\widetilde{K}_i}(\Omega,{\mathscr G}_L)$.
Then, one can give $C^\infty_0(\Omega,{\mathscr G}_L)$ a limit topology induced by finite
family of norms~\cite{NK}
\begin{align}
\|\widetilde{\phi}\|_{\widetilde{K},m}=\sup_{\stackrel{|p|\leq m}{x \in \widetilde{K}}}
|D^p \widetilde{\phi}(x)|=\sup_{\stackrel{|p|\leq m}{x \in \widetilde{K}}} 
\left\{\sum_{(\mu_1,\ldots,\mu_k)\in M_L^0}
|D^p \widetilde{\phi}_{\mu_1,\ldots,\mu_k}(x)|\right\}\,\,.
\label{snorma} 
\end{align}

Finally, a suitable topological structure to the space $G_0^\infty(U,{\mathscr G}_L)$ of
${\mathscr G}_L$-valued superfunctions on an open set $U \subset {\mathscr G}_L^{m,n}$
which have compact support, it is obtained immediately by the natural identification of
${\mathscr G}_L^{m,n}$ with ${\Bbb R}^{2^{L-1}(m+n)}$ and by the obvious extension
of the construction above, which allows us define a limit topology induced to
the space $G_0^\infty(U,{\mathscr G}_L)$ by finite family of norms,
\begin{align}
\|\lambda(\phi)\|_{K,m+n}=\sup_{\stackrel{|p|\leq m+n}{z \in K}}
|D^p (\lambda(\phi))(z)|\,\,,
\qquad D^p=\frac{\partial^{|q|+|r|}}{\partial x^{q_1}_1 \cdots \partial x^{q_m}_m
\partial\theta^{r_1}_1 \cdots \partial\theta^{r_n}_n}\,\, 
\label{snorm3} 
\end{align}
where the derivatives $\partial^{|q|}/\partial x^{q_1}_1 \cdots \partial x^{q_m}_m$ commute
while the derivatives $\partial^{|r|}/\partial\theta^{r_1}_1\cdots \partial\theta^{r_n}_n$
anticommute, and $|p|=|q|+|r|=\sum_{i=1}^m q_i+\sum_{j=1}^n r_j$ defines the total order of
the derivative, with $r_j=0,1$.

We are now ready to define a superdistribution in an open subset $U$ of
${\mathscr G}_L^{m,n}$. The set of all superdistributions in $U$ will be
denoted by ${\mathfrak D}^\prime(U)$. A superdistribution is a continuous linear functional
$u:G_0^\infty(U)\rightarrow {\mathscr G}_L$, where $G_0^\infty(U)$ denotes
the test superfunction space of $G^\infty(U)$ superfunctions with compact
support in $K \subset U$. The continuity of $u$ on 
$G_0^\infty(U)$ is equivalent to its boundedness on a
neighbourhood of zero, i.e., the set of numbers $u(\phi)$ is bounded for all
$\phi \in G_0^\infty(U)$. The last statement translates directly into:

\begin{proposition}
A superdistribution $u$ in $U \in {\mathscr G}_{L}^{m,n}$ is a continuous linear functional on
$G^{\infty}_{0}(U)$ if and only if to every compact set $K \subset U$, there exists a constant
$C$ and $(m+n)$ such that 
\[
\left|u(\phi)\right| \leq C 
\sup_{\stackrel{{|p| \leq m+n}}{z \in K}} 
\left|D^{p}(\phi)(z)\right|\,\,, 
\quad \phi \in G^{\infty}_{0}(K)\,\,.
\]
\label{dspdis} 
\end{proposition}

\begin{proof}
See~\cite{DanCaio}
\end{proof}

\section{Wavefront Set of a Superdistribution}
A great deal of progress has been made in recent years in characterizing the
``ultraviolet divergences'' of quantum fields in curved spacetime and developing
renormalization theory for interacting quantum fields by the use of the methods
of ``microlocal analysis.'' This leads to the definition of the 
wavefront set, denoted (${{WF}}$), of a distribution, a refined description of 
the singularity spectrum. Similar notion was developed in other versions by 
Sato~\cite{Sa}, Iagolnitzer~\cite{Ia} and Sj\"ostrand~\cite{Sj}. The 
definition as known nowadays is due to H\"ormander. He used this 
terminology due to an existing analogy between his studies on the 
``propagation'' of singularities and the classical construction of 
propagating waves by Huyghens. For a distribution $u$ we introduce its wavefront set
${{WF}}(u)$ as a subset in phase space ${\Bbb R}^n\times{\Bbb R}^n$.\footnote{The
functorially correct definition of phase space is ${\Bbb R}^n\times({\Bbb R}^n)^*$. We shall
here ignore any attempt to distinguish between ${\Bbb R}^n$ and $({\Bbb R}^n)^*$.}
We shall be thinking of points $(x,k)$ in phase space as specifying those singular directions
$k$ of a ``bad'' behaviour of the Fourier transform $\widehat{u}$ at infinity that are
responsible for the non-smoothness of $u$ at the point $x$ in position space. So we shall
usually want $k \not= 0$. A relevant point is that ${{WF}}(u)$ is independent of the coordinate
system chosen, and it can be described locally.

It is well-known that the regularity properties of a distribution are in cor\-res\-pondence
with the decay properties of its Fourier transform (see Appendix A for details). The
results which now follow prove that the decay properties of a superdistribution
at infinity and the smoothness properties of its Fourier transform are analogous to the
case of ordinary distributions, i.e., no new singularity appear by taking into account
the structure of the superspace.

\begin{lemma} Let $X \subset {\mathscr G}_L^{m,0}$ be an open set, and
$u$ be a superdistribution on $X$ taking values in ${\mathscr G}_L$, i.e., a
linear functional $u:G_0^\infty(X) \rightarrow {\mathscr G}_L$. Let $\phi$ be a
supersmooth function with compact support $K \subset X$. Then $\phi u$ is
also supersmooth on $K$, if its components $(\phi u)(\epsilon(x))$ are smooth on a compact
set $K^\prime \subset \Omega$, where $\Omega$ is the body of superspace. Therefore, the
following estimate holds:
\begin{equation*}
\left| \widehat{\phi u}(k) \right| 
\leq (1+|k_{\bf b}|)^{-N}C(N,\phi)\,\,.
\end{equation*}
\label{main}
\end{lemma}

\begin{proof}
See~\cite{DanCaio}
\end{proof}

\begin{lemma} By replacing ${\mathscr G}_L^{m,0}$ by ${\mathscr G}_L^{m,n}$
in the Lemma \ref{main}, then the following estimate holds:
\begin{align*}
\left| \widehat{\phi u}(k,\theta,\bar{\theta}) \right| 
\leq (1+|k_{\bf b}|)^{-N}C(N,\phi_{(\gamma)})
\|\theta_1\|\|\bar{\theta}_1\|\cdots \|\theta_n\|\|\bar{\theta}_n\|\,\,.
\end{align*}
\end{lemma}

\begin{proof}
See~\cite{DanCaio}
\end{proof}

Combining the results above, we have proved:

\begin{theorem}  The singularities of a superdistribution $u$ are located at specific values 
of the body of $x$, the coordinates of the {\bf physical spacetime}, independently of the odd
coordinates.
\label{prop1}
\end{theorem}

We sum up the preceding discussion as follows:

\begin{definition}[Wavefront Set of a Superdistribution] The wavefront set $WF(u)$ of a
superdistribution $u$ in a superspace ${\mathscr M}$ is the complement of the set of
all regular directed points in the cotangent bundle $T^*{\mathscr M}_0$,
where ${\mathscr M}_0=\epsilon({\mathscr M})$ is the body of superspace,
excluding the trivial point $k_{\bf b}=0$.
\label{mscs0}
\end{definition}

\begin{remark}
A direction $k_{\bf b}$ for which the Fourier transform of a superdistribution 
$u$ shows to be of fast decrease is called to be a 
{\em regular direction} of $\hat{u}$.
Therefore, in order to determine whether $(x_{\bf b},k_{\bf b})$ belongs to the wavefront
set of $u$ one must first to localize $u$ around $x_{\bf b}$, next to obtain Fourier
transform $\hat{u}$ and finally to look at the decay in the direction
$k_{\bf b}$. Hence, the wavefront set not only describes the set of points where a
superdistribution is singular, but it also localizes the frequencies that constitute
these singularities.
\end{remark}

There is a more precise version of Definition \ref{mscs0}. As we have seen in
Section \ref{DiSmf} all of the foregoing definitions and statements about supermanifolds
may be converted into corresponding definitions and statements about ordinary manifolds,
since associated with a supermanifold ${\mathscr M}$ of dimension $(m,n)$ is a family
of ordinary manifolds, of dimensions $N=2^{L-1}(m+n)$, $(L=1,2,\ldots)$. The resulting
manifold is called the $L$th skeleton of $\mathscr M$ and denoted by
${\mathscr S}_L({\mathscr M})$~\cite{DeWitt}. With the aid of the family of skeletons we can
define the pushforward (or direct image) of a superdistribution. Let
$X \subset {\mathscr S}_L({\mathscr M})$ and $Y \subset {\mathscr M}_0$ be open sets and let
$\epsilon$ be the natural projection from ${\mathscr S}_L({\mathscr M})$ (or ${\mathscr M}$) to
${\mathscr M}_0$, the body map. If we introduce local coordinates $x=(x_1,\ldots,x_N)$ in
$X$, then $Y$ is defined by $x_{\bf b}=(x_1,\ldots,x_m)$. There is a local relationship
between the body and the skeletons given by
\[
{\mathscr S}_L(X)\overset{\text{diff.}}{=}Y \times {\Bbb R}^{2^{L-1}(m+n)-m}\,\,.
\]
Now, let $u$ be a superdistribution on $X$, then the pushforward $\epsilon_*u$ defined by
$\epsilon_*u(\varphi)=u(\epsilon^*\varphi)$, $\varphi \in C_0^\infty(Y)$, it is a
superdistribution on $Y$. Using these concepts, we can establish the following

\begin{corollary}
Let $\epsilon:X \subset{\mathscr S}_L({\mathscr M})\rightarrow Y \subset{\mathscr M}_0$ be the
body projection, and let $u \in {\mathfrak D}^\prime(X)$. Then
\begin{align*}
WF(\epsilon_*u)\subset \Bigl\{(x_{\bf b},k_{\bf b}) \in T^*{\mathscr M}_0\backslash 0
\mid\,\exists\,\,x^\prime=(x_{m+1},\ldots,x_{N^\prime}),
(x_{\bf b},x^\prime,k_{\bf b},0)\in WF(u)\Bigr\}\,,
\end{align*}
where $N^\prime=2^{L-1}(m+n)-m$.
\end{corollary}

\begin{proof}
See~\cite{DanCaio}
\end{proof}

\section{Conclusions}
We have introduced a notion of superdistribution in superspace which seems to have
some advantages: by exploring the functorial relations between a $G^\infty$-superspace
and a family of Banach manifolds $C^\infty$ we define the space of superdistributions
as the dual of the test function space of $C^\infty$-functions with compact support
endowed with a suitable topology on Banach spaces. In particular, Wightman superfunctions
and superpropagators, which appear in the supersymmetric quantum field theory, can be
treated as our superdistributions. We have also obtained useful results on the
sin\-gu\-la\-ri\-ty structure of such objects, here analysed in the context of
the development of the potent mathematical tool of microlocal analysis and
characterized in terms of the its wavefront set. Our analysis represents only
the first step towards a more interesting physically situation: the perturbative
treatment of interacting quantum superfield models, in particular the formulation
of renormalization theory on curved supermanifolds.

\section*{Acknowledgments.}
We thank C.M.M. Polito for his collaboration in the later stages of this work,
when ours results were extended for a general supermanifold.

\appendix 
\renewcommand{\theequation}{\Alph{section}.\arabic{equation}}
\renewcommand{\thesection}{\Alph{section}}

\setcounter{equation}{0} \setcounter{section}{0}

\section{Microlocal Analysis: Review of Some Basic Ideas}
In this appendix we briefly recall some standard facts on microlocal analysis.
The key point of the microlocal analysis is the transference of 
the study of singularities of distributions from the configuration 
space only to the rather phase space, by exploring in frequency space the decay properties
of a distribution at infinity and the smoothness properties of its 
Fourier transform. As it is well-known~\cite{RS2, Hor2}, a distribution of compact 
support, $u \in {\mathscr E}^\prime({\Bbb R}^n)$, 
is a smooth function if, and only if, its 
Fourier transform, $\widehat{u}$, rapidly decreases at infinity
(i.e., as long as supp$\,u$ does not touch the singularity points). By a fast 
decay at infinity, one must understanding that for all positive integer $N$ 
exists a constant $C_N$, which depends on $N$, such that 
\begin{equation}
|\widehat{u}(k)| \leq (1+|k|)^{-N}C_N\,, 
\qquad \forall\,N \in {\Bbb N};\,\,k \in {\Bbb R}^n\,\,. 
\label{Pepe} 
\end{equation}
If, however, $u \in {\mathscr E}^\prime({\Bbb R}^n)$ is not smooth, then the 
directions along which $\widehat{u}$ does not fall off sufficiently fast 
may be adopted to characterize the singularities of $u$.

For a distribution does not necessarily of compact support, still we can verify if its Fourier
transform rapidly decreases in a given region $V$ through the technique of localization.
More precisely, if $V \subset X \subset {\Bbb R}^n$ and $u \in {\mathscr D}^\prime(X)$,
we can restrict $u$ to a distribution $u|_V$ in $V$ by setting $u|_V(\phi)=u(\phi)$, where
$\phi$ is a smooth function with support contained in a region $V$. The distribution $\phi u$
can then be seen as a distribution of compact
support on ${\Bbb R}^n$. Its Fourier transform will be defined as a distribution on ${\Bbb R}^n$,
and must satisfy, in absence of singularities in $V \subset {\Bbb R}^n$, the property (\ref{Pepe}).
From this point of view, all development is local in the sense that only the behaviour of the
distribution on the arbitrarily small neighbourhood of the singular point, in the configuration
space, is relevant.
 
Let $u \in {\mathscr D}^\prime({\Bbb R}^n)$ be a distribution and $\phi \in 
C_0^\infty(V)$ a smooth function with support $V \subset {\Bbb R}^n$. Then, 
$\phi u$ has compact support. The Fourier transform of $\phi u$ produces a smooth function in
frequency space. 

\begin{lemma}
Consider $u \in {\mathscr D}^\prime({\Bbb R}^n)$ and $\phi \in C_0^\infty(V)$. Then
$\widehat{\phi u}(k)=u(\phi e^{-ikx})$. Moreover, the restriction of $u$ to
$V \subset {\Bbb R}^n$ is smooth on $V$ if, and only if, for every $\phi \in C_0^\infty(V)$ 
and each positive integer $N$ there exist a constant $C(\phi, N)$, which depends on $N$ and
$\phi$, such that $|\widehat{\phi u}(k)| \leq (1+|k|)^{-N}C(\phi, N)$, for all $N \in {\Bbb N}$
and $k \in {\Bbb R}^n$.
\label{lema1}
\end{lemma}

If $u \in {\mathscr D}^\prime({\Bbb R}^n)$ is singular in $x$, and $\phi \in 
C_0^\infty(V)$ is $\phi(x)\not= 0$; then $\phi u$ is also singular 
in $x$ and has compact support. However, in some directions in $k$-space 
$\widehat{\phi u}$ until will be asymptotically limited. 
This is called the set of {\em regular directions} of $u$. 

\begin{definition} Let $u(x)$ be an arbitrary distribution,
not necessarily of compact support, on an open set $X \subset {\Bbb R}^{n}$. Then, the set of
pairs composed by singular points $x$ in configuration space and by its associated
nonzero singular directions $k$ in Fourier space 
\begin{align}
{{WF}}(u)=\{(x,k) \in X \times ({\Bbb R}^n\backslash 0)\left|\right. k 
\in \Sigma_x(u)\}\,\,, 
\label{A.2}
\end{align}
is called {\bf wavefront set} of $u$. $\Sigma_x(u)$ is defined to be the 
complement in ${\Bbb R}^n\backslash 0$ of the set of all $k \in {\Bbb 
R}^n\backslash 0$ for which there is an open conic neighbourhood $M$ of $k$ 
such that $\widehat{\phi u}$ rapidly decreases in $M$, for $|k| \rightarrow \infty$. 
\end{definition}

\begin{remarks}
We will now collect some basic properties of the wavefront set:
\begin{enumerate}

\item The ${{WF}}(u)$ is conic in the sense that it remains invariant under the action of 
dilatations, i.e., when we multiply the second variable by a positive 
scalar. This means that if $(x,k) \in {{WF}}(u)$ then
$(x,\lambda k)\in {{WF}}(u)$ for all $\lambda > 0$. 

\item  From the definition of ${{WF}}(u)$, it follows that $\pi_1({{WF}}(u)) \rightarrow x$
is the projection onto the first variable, by consisting of those points that have no
neighbourhood wherein $u$ is a smooth function. The projection onto the second variable,
$\pi_2({{WF}}(u)) \rightarrow \Sigma_x(u)$, is the cone around $k$ attached to a such point
de\-no\-ting the set of high-frequency directions responsible for the appearance of a
singularity at this point.

\item The wavefront set of a smooth function is the empty set.

\item For all smooth function $\phi$ with compact suport ${WF}(\phi u)\subset {WF}(u)$.

\item For any partial linear differential operator $P$, with $C^\infty$ coefficients, we have
\[
{WF}(Pu)\subseteq {WF}(u)\,\,.
\]

\item If $u$ and $v$ are two distributions belonging to ${\mathscr D}^\prime({\Bbb R}^n)$,
with wavefront sets ${WF}(u)$ and ${WF}(v)$, respectively; then the wavefront set of
$(u+v) \in {\mathscr D}^\prime({\Bbb R}^n)$ is contained in ${WF}(u)\cup{WF}(v)$.

\item If $U,V$ are open set of ${\Bbb R}^n$, $u \in {\mathscr D}^\prime(V)$, and
$\chi:U \rightarrow V$ a diffeomorphism such that $\chi^*u \in {\mathscr D}^\prime(U)$ is
the distribution pulled back by $\chi$, then ${WF}(\chi^*u)=\chi^*{WF}(u)$.

\end{enumerate}
\end{remarks}

We emphasize that a number of operations, not possible in general, 
become feasible for distributions under special assumptions on 
their wavefront set, such as taking products. As a result of this, the wavefront set applies
to theories which are formulated in terms of pointlike fields. In the naive 
perturbative scheme of quantum field theories, one encounters formal 
products of fields which are a priori ill-defined. This 
difficulty lies at the heart of renormalization theory. The latter starts 
from the observation that products of fields (operator-valued distributions) 
are well-defined on a subset which does not contain the diagonal (all 
coinciding points, or the zero section). Renormalization consists then in 
the continuation of products of distributions to the whole space. 

In order to give precise statements to the product of these fields,
we appeal to the criterion below:

\begin{theorem}[H\"ormander's Criterion]
Let $u$ and $v$ be distributions;
if the wavefront set of $u$ and $v$ are such that
\[  
(x,0)\not\in {{WF}}(u) \oplus {{WF}}(v)=
\{(x,k_1+k_2)\left.\right|(x,k_1)\in {{WF}}(u),
(x,k_2)\in {{WF}}(v)\}\,\,,
\]
then the product $uv$ exists and 
${{WF}}(uv)\subset{{WF}}(u)\cup{{WF}}(v)\cup
({{WF}}(u)\oplus{{WF}}(v))$. 
\end{theorem}

Hence, the product of the distributions $u$ and $v$ is well-defined 
in $x$, if $u$, or $v$, or both distributions are regular in $x$. 
Otherwise, if $u$ and $v$ are singular in $x$, the product can still exist 
if, the sum of the second components of ${{WF}}(u)$ and ${{WF}}(v)$  
related to $x$ can be linearly combined to give zero only by a trivial solution.

\begin{example}
The distributions $u,v \in {\mathscr D}^\prime({\Bbb R})$, 
$u(x)=\frac{1}{x+i\epsilon}$ and $v(x)=\frac{1}{x-i\epsilon}$, with the
Heavyside distributions $\widehat{u}(k)=2\pi i\theta(-k)$ and 
$\widehat{v}(k)=-2\pi i\theta(k)$ as their Fourier transforms, have the
following wavefront sets:    
\begin{align*}
{WF}(u)=\{(0,k) \mid k \in {\Bbb R}^-\backslash 0\}\,\,,\quad
{WF}(v)=\{(0,k) \mid k \in {\Bbb R}^+\backslash 0\}\,\,.
\end{align*}
Thus, from the H\"or\-man\-der's Criterion one finds that there exist the powers 
of $u^n$ and $v^n$. On the other hand, the product between $u$ and $v$ do not match the 
above criterion and do not exist, indeed. The example clearly indicates that one 
can multiply distributions even if they have overlapping singularities, provided 
their wavefront sets are in favorable positions. Such an observation is
significant because it makes clear that {\em the problem is not only where the
support is, but in which directions the Fourier transform is not rapidly
decreasing}!
\end{example}

Another result, which we merely state, is needed to complete this briefing
on microlocal analysis.

\begin{theorem}[Wavefront set of pushforwards of a distribution] Let
$f:X \rightarrow Y$ be
a submersion, and let $u \in {\mathscr E}^\prime(X)$. Then
\[
WF(f_*u) \subset \{(f(x),\eta) \mid x \in X, (x,^t\!\!f_x^\prime \eta)\in
WF(u)\,\,{\mbox{or}}\,\,^t\!f_x^\prime \eta=0\}\,\,,
\]
where $^t\!f_x^\prime$ denotes the transpose matrix of the Jacobian matrix $f_x^\prime$ of
$f$.
\label{wfspd}
\end{theorem}


\end{document}